\documentclass{emulateapj}

\usepackage{graphicx}
\usepackage{epsfig}
\usepackage{natbib}
\usepackage{multirow}
\usepackage{amsmath,amssymb}
\usepackage{rotating}
\usepackage{booktabs}

\begin{document}
\submitted{ApJ in press}
\journalinfo{Accepted for publication in The Astrophysical Journal}

\title{Resolving the Cosmic Far-Infrared Background at 450 and 850 Microns with SCUBA-2}
\author{Chian-Chou Chen \altaffilmark{1}, Lennox L. Cowie\altaffilmark{1}, Amy J. Barger\altaffilmark{1,2,3},  Caitlin. M. Casey\altaffilmark{1,$\star$}, Nicholas Lee\altaffilmark{1}, David B. Sanders\altaffilmark{1}, Wei-Hao Wang \altaffilmark{4}, Jonathan P. Williams\altaffilmark{1}}
\altaffiltext{1}{Institute for Astronomy, University of Hawaii, 2680 Woodlawn Drive, Honolulu, HI 96822.}
\altaffiltext{2}{Department of Astronomy, University of Wisconsin-Madison, 475 North Charter Street, Madison, WI 53706.}
\altaffiltext{3}{Department of Physics and Astronomy, University of Hawaii, 2505 Correa Road, Honolulu, HI 96822.}
\altaffiltext{4}{Academia Sinica Institute of Astronomy and Astrophysics, P.O. Box 23-141, Taipei 10617, Taiwan.}
\altaffiltext{$\star$}{Hubble Fellow}
\subjectheadings{cosmology: observations|  galaxies: evolution  |  galaxies: formation  |  submillimeter}

\begin{abstract}
We use the SCUBA-2 submillimeter camera mounted on the JCMT to obtain extremely deep number counts at 450 and 850\,$\mu$m. We combine data on two cluster lensing fields, A1689 and A370, and three blank fields, CDF-N, CDF-S, and COSMOS, to measure the counts over a wide flux range at each wavelength. We use statistical fits
to broken power law representations to determine the number counts. 
This allows us to probe to the deepest possible level in the data. 
At both wavelengths our results agree well with the literature in the flux range over which they have been measured, with the exception of the 850\,$\mu$m counts in CDF-S, where we do not observe the counts deficit found by previous single-dish observations. At 450\,$\mu$m, we detect significant counts down to $\sim$1\,mJy, an unprecedented depth at this wavelength. By integrating the number counts above this flux limit, we measure
$113.9^{+49.7}_{-28.4}$\,Jy\,deg$^{-2}$ of the 450\,$\mu$m extragalactic background light (EBL). The majority of this contribution is from sources with S$_{450~\mu{\rm m}}$ between 1--10\,mJy, and these sources are likely to be the ones that are analogous to the local luminous infrared galaxies (LIRGs). At 850\,$\mu$m, we measure $37.3^{+21.1}_{-12.9}$\,Jy\,deg$^{-2}$ of the EBL. Because of the large systematic uncertainties on the COBE measurements, the percentage of the EBL we resolve could range from 48--153\% (44--178\%) at 450 (850)\,$\mu$m. Based on high-resolution SMA observations of around half of the 4\,$\sigma$ 850\,$\mu$m sample in CDF-N, we find that  { 12.5$^{ +12.1}_{ -6.8}$\%} of the sources are blends of multiple fainter sources. This is a low multiple fraction, and we find no significant difference between our original SCUBA-2 850\,$\mu$m counts and the multiplicity corrected counts.
\end{abstract}

\section{Introduction}
Following the discovery of the far-infrared (FIR) 
extragalactic background light (EBL) by the COBE satellite \citep{Puget:1996p2082,Fixsen:1998p2076}, many surveys have been conducted to detect the sources producing this light. Such studies have used both ground-based telescopes (e.g., \citealt{Smail:1997p6820, Barger:1998p13566, Hughes:1998p9666}) and space-based satellite missions (e.g., \citealt{Oliver:2010p11204, Berta:2011yq}).
Given that there is a comparable amount of light absorbed by dust and re-radiated in the FIR as there is seen directly in the UV/optical \citep{Dole:2006p9898}, the dusty sources uncovered by these surveys are key in the development of a full understanding of galaxy formation. The FIR number counts provide simple yet fundamental constraints on empirical models (e.g., \citealt{Valiante:2009rt, Bethermin:2011vn}) and semi-analytical simulations \citep{Hayward:2013qy,Hayward:2013lr}.  

The construction of the FIR number counts began with 850/450\,$\mu$m observations made using the SCUBA camera \citep{Holland:1999fk} mounted on the 15-meter James Clerk Maxwell Telescope (JCMT) \citep{Smail:1997p6820, Barger:1998p13566, Barger:1999p6485,Hughes:1998p9666,Eales:1999p9715,Eales:2000uq,Cowie:2002p2075, Scott:2002p6539, Smail:2002p6793, Borys:2003p6612, Serjeant:2003lr, Webb:2003p6591, Wang:2004p2270, Coppin:2006p9123, Knudsen:2008p3824, Zemcov:2010uq}. Since then, many single-dish telescopes, instruments, and missions have been developed to survey the sky at FIR through millimeter wavelengths. At $\lambda < 500\,\mu{\rm m}$, the number counts have been established by the second-generation Submillimeter High Angular Resolution Camera (SHARC-2;  \citealt{Dowell:2003yq}) at the Caltech Submillimeter Observatory (CSO) (e.g., \citealt{Khan:2007vn}), the Balloon-borne Large Aperture Submillimeter Telescope (BLAST; \citealt{Pascale:2008fj})(e.g., \citealt{Devlin:2009kx}),  the {\it Herschel Space Observatory} (hereafter {\it Herschel}; \citealt{Pilbratt:2010lr})(e.g., \citealt{Oliver:2010p11204, Berta:2011yq}). At $\lambda > 500\,\mu {\rm m}$, in addition to SCUBA, the number counts have been probed by the LABOCA camera \citep{Siringo:2009rt} on the Atacama Pathfinder Experiment (APEX; \citealt{Gusten:2006mz})(e.g., \citealt{Weis:2009qy}), the AzTEC \citep{Wilson:2008ys} camera on both the JCMT (e.g., \citealt{Perera:2008th, Austermann:2010ly}) and the Atacama Submillimeter Telescope Experiment (ASTE; \citealt{Ezawa:2004qy})(e.g., \citealt{Aretxaga:2011op, Scott:2010pp, Scott:2012mn}), the Max-Planck Bolometer array (MAMBO; \citealt{Kreysa:1998fr}) on the IRAM 30 m telescope (e.g., \citealt{Greve:2004p6618, Bertoldi:2007gf}), and Bolocam \citep{Glenn:1998zr} on the CSO (e.g., \citealt{Laurent:2005ve}).

The biggest challenge for constructing the number counts in the FIR is poor spatial resolution (typically $>$\,10$''$), due to the diffraction limits of the single-dish telescopes at longer wavelengths. Poor resolution has imposed a fundamental limitation, the confusion limit \citep{Condon:1974qy}, on our ability to resolve directly the faint sources that contribute the bulk of the background light. Poor resolution also prevents us from resolving close pairs within the large beam sizes. 
Interferometric observations \citep{Wang:2011p9293, Barger:2012lr, Smolcic:2012pp,Karim:2013fk,Hodge:2013lr} and semi-analytical models \citep{Hayward:2013qy, Hayward:2013lr} have shown these to be common, and their effects must be understood in order to construct the true counts.

Techniques have been developed to work around the problem of the confusion limit. Surveys targeting massive galaxy clusters have unveiled a few faint sources with fluxes many times below the confusion limit through gravitational magnification \citep{Smail:1997p6820,Smail:2002p6793, Cowie:2002p2075, Knudsen:2008p3824, Johansson:2011zr, Chen:2013fk}, though the positional uncertainties can still cause large uncertainties in the amplifications and in the intrinsic source fluxes (\citealt{Chen:2011p11605}). In blank-field surveys, probability of deflection analyses, or P(D), using the number distribution of pixel values put stringent constraints on counts deeper than the confusion limit \citep{Scheuer:1957fk,Weis:2009qy,Scott:2010pp,Glenn:2010kx,Lindner:2011ys}. 

The new SCUBA-2 camera \citep{Holland:2013lr} mounted on the JCMT provides the fastest mapping capability at 450 and 850\,$\mu$m with the best FIR spatial resolution (FWHM $\sim$ 7$\farcs$5) at 450\,$\mu$m among single-dish FIR telescopes. This greatly enhances our ability to resolve the 450\,$\mu$m EBL, thanks to the smaller confusion limit, and makes the 450\,$\mu$m number counts less affected by close pairs. In addition, the better positions provide better determinations of the lensing amplifications.  

We have used SCUBA-2 to target two well-studied massive lensing clusters, A370 and A1689, and three blank fields, COSMOS, CDF-N, and CDF-S, in order to construct the 450\,$\mu$m number counts over the widest possible flux range. We have previously shown some of these results on A370 in \citet{Chen:2013fk} and on COSMOS in \citet{Casey:2013uq}. In this paper, we present the full results from all five fields, and we combine the data to measure the number counts.  The details of the observations and the data reduction are presented in Section~2. In Section~3, we explain our methodology for constructing the number counts, which uses a combination of gravitational lensing and P(D) analysis, and present our results. We discuss the effects of field-to-field variance and source
blending (multiplicity) in Section~4. We discuss the implications of our results in Section~\ref{secdisc}, and we provide a brief summary in Section~\ref{secsum}.

\section{Observations and Data Reduction}
The SCUBA-2 data were taken  between late 2011 and early 2013. The observations on A1689, A370, and COSMOS were taken under the best weather conditions (band 1, $\tau_{\rm 225~GHz}<0.05$). Most of the observations carried out on CDF-N and CDF-S were under band 2 ($ 0.05 < \tau_{\rm 225~GHz}<0.08$)\footnote{The program IDs are M11BH15B, M11BH11A, M11BH26A, M12AH15B, M12AH11A, M12AH26A, M12BH34B, M12BH26A, M12BH21A, M13AH29A, and M13AH24A.}. To cover a wide range of submillimeter fluxes, we used the \textsc{CV Daisy} scan pattern on both the massive cluster fields, where the smaller field is well matched to the strong lensing regions in the clusters, and the blank fields (except COSMOS). In order to cover larger areas uniformly to find brighter but rarer sources, we also used the \textsc{PONG-900} scan pattern on the blank fields. Detailed information about the SCUBA-2 scan patterns can be found in \citet{Holland:2013lr}. We summarize the details of our observations in Table \ref{obs}. 

\begin{table*}
\begin{center}
\caption{SCUBA-2 Observations}
\begin{tabular}{cccccccc}
\hline
\hline
 \multirow{3}{*}{Field} & \multicolumn{2}{c}{Centroid Coordinate} &\multirow{3}{*}{Weather}   &  \multirow{3}{*}{Scan Mode}  &\multirow{2}{*}{Total Exposure}&  Effective Area$^a$  &$\bar{\sigma}^b$    \\
 &R.A.(J2000)&Decl.(J2000)&&& &[450\,$\mu$m, 850\,$\mu$m] &[450\,$\mu$m, 850\,$\mu$m] \\
 &(H:M:S)&(D:M:S)&&&(hours)&(arcmin$^2$)&(mJy/Beam)\\
\hline
 A1689   &13:11:29.8&$-$01:20:35.8&     1  &    \textsc{CV Daisy}  &14.8&[69.4, 73.0] &      [ 4.5, 0.79]  \\
CDF-N    &12:36:49.6&+62:13:53.0 & 1+2  &    \textsc{CV Daisy} + \textsc{PONG-900}  &24.7&[102.6, 104.7] &      [ 8.8, 0.83]	    \\
CDF-S	&03:32:28.0&$-$27:48:30.0&2	&     \textsc{CV Daisy} + \textsc{PONG-900} & 22.9& [97.0, 102.5] & [10.3, 0.88] \\
A370        &02:39:53.0&$-$01:34:38.0&  1  &    \textsc{CV Daisy}  &13.7&[83.1, 83.9]&      [ 5.3, 1.07]   \\
COSMOS    &10:00:24.0&+02:24:00.0 &   1  &    \textsc{PONG-900}  &38.0&[286.0, 281.7]&      [ 4.7,  0.87]  \\
         
\hline
&&&&&&& \\
 \multicolumn{8}{l}{$^{a}$ Total area to 1.5 -- 2 times the central noise level. In cluster fields they represent the total source plane area.} \\ 
 \multicolumn{8}{l}{$^{b}$ Average 1\,$\sigma$ sensitivity within the effective area. Note that the quoted values for A370 and COSMOS are slightly higher than those} \\
 \multicolumn{8}{l}{\ \ quoted by \citet{Chen:2013fk} and \citet{Casey:2013uq}, as they were quoting maximum sensitivities.} \\ 
\end{tabular}
\label{obs}
\end{center}
\end{table*}

We reduced the data using the Dynamic Iterative Map Maker (DIMM) in the SMURF package from the STARLINK software developed by the Joint Astronomy Centre (\citealt{Jenness:2011lr, Chapin:2013fk}) released after July 1, 2012. DIMM performs iterative estimations on the common mode signal, the astronomical signal, and the white noise. It also does flatfield and extinction corrections and applies a Fourier Transform filter to remove low-frequency excess signal relative to the white noise that is not able to be removed through common mode subtraction (\citealt{Chapin:2013fk}). We adopted the standard configuration file \textit{dimmconfig\_blank\_field.lis} for our science purposes. We ran DIMM on each bolometer subarray individually to avoid data splitting, and we used the \textsc{MOSAIC\_JCMT\_IMAGES} recipe in PICARD, Pipeline for Combining and Analyzing Reduced Data (\citealt{Jenness:2008fk}), to coadd the products into final maps. 

To increase the detectability of point sources, as nearly all the SMGs are expected to be much smaller than the $\sim$\,10$''$ resolution, we applied a matched-filter to our maps. This provides a maximum likelihood estimate of the source strength (e.g., \citealt{Serjeant:2003lr}). Assuming S(i,j) and $\sigma$(i,j) are the signal and r.m.s noise maps produced by DIMM, and PSF(i,j) is the signal point spread function, the filtered signal map F(i,j) would be 
\begin{equation}
F(i,j) = \frac{\sum_{i,j} [S(i,j) / \sigma(i,j)^2 \times PSF(i,j)]}{\sum_{i,j} [1 / \sigma(i,j)^2 \times PSF(i,j)^2]}, 
\end{equation} 
and the filtered noise map N(i,j) would be 
\begin{equation}
N(i,j) = \frac{1}{\sqrt{\sum_{i,j} [1 / \sigma(i,j)^2 \times PSF(i,j)^2]}}. 
\end{equation} 

Ideally, the PSF for the matched-filter algorithm is a Gaussian normalized to a peak of unity with FWHM equal to the JCMT beam size at a given wavelength (i.e., $7\farcs5$ at 450\,$\mu$m and $14''$ at 850\,$\mu$m). However, the map produced from DIMM usually has low spatial frequency structures that need to be subtracted off before performing the source extraction. Thus, before running the matched-filter, we convolved the map with a broad Gaussian normalized to a sum of unity, and we subtracted this convolved map from the original map. Note that in \citet{Chen:2013fk}, we showed that the source fluxes and the S/N are not sensitive to the size of the FWHM for reasonable choices. Thus, we simply adopted the default values ($20''$ at 450\,$\mu$m and $30''$ at 850\,$\mu$m). To optimize the S/N, we processed the PSF used for the matched-filter similarly. It becomes a Gaussian with a convolved broader Gaussian subtracted off, which gives a Mexican hat-like wavelet. We adopted the PICARD recipe SCUBA2\_MATCHED\_FILTER for the tasks described above. 

\begin{figure}
 \begin{center}
    \leavevmode
      \includegraphics[scale=0.6]{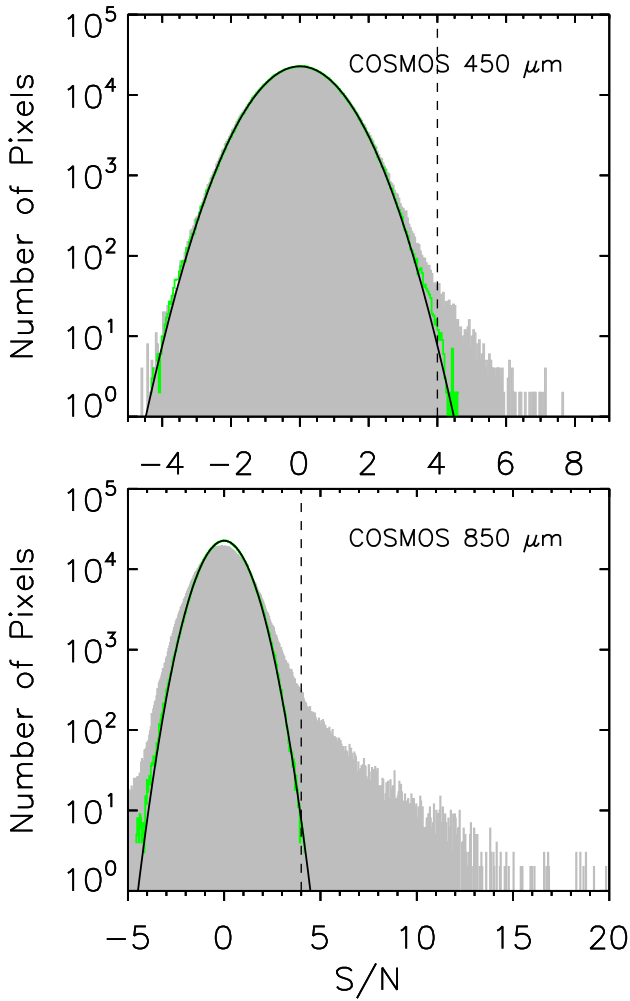}
       \caption{The 450~$\mu$m and 850~$\mu$m S/N histograms for the 
pixels located within the regions of the COSMOS signal map 
where the noise level is less than 1.5 times the central noise level {\it (gray shading)\/} 
and for the pixels in the corresponding regions of the true noise map {\it (green)}.
The black curves are the expected pure noise distributions with $\sigma=1$. 
Dashed vertical lines are 4\,$\sigma$ cuts. 
}
     \label{histo}
  \end{center}
\end{figure}

We then calibrated the fluxes using standard Flux Conversion Factors (FCFs; 491\,Jy\,pW$^{-1}$ for 450\,$\mu$m and 537\,Jy\,pW$^{-1}$ for 850\,$\mu$m). The relative calibration accuracy is shown to be stable and good to 10\% at 450\,$\mu$m and 5\% at 850\,$\mu$m (\citealt{Dempsey:2013qy}). Ten per cent upward corrections were applied to compensate for the flux lost during filtering, which we estimated from simulations using fake sources. We also tested the accuracy of the flux calibrations by---instead of using the standard FCFs---adopting the FCFs obtained from the calibrators in each night of observations, as was done in \citet{Chen:2013fk}. We found the results agree to better than 10\%, which is essentially the uncertainty of the FCFs.
 
\section{Number Counts}
In order to measure the galaxy number counts, we need source-free maps with only pure noise to estimate how many fake sources are contaminating the counts. We followed the procedure in \citet{Chen:2013fk}. For each wavelength, we generated two data maps, each with roughly half of the total exposure time, and we subtracted them to obtain the source-free maps. We rescaled the value of each pixel following the equation $\sqrt{t1\times t2}$/(t1+t2), with t1 and t2 representing the exposure time of each pixel from the two maps. Finally, we applied the matched filter and FCFs, as we did on the signal maps. We produced these true noise maps, sometimes referred to as jackknife maps in the literature, for each of our fields. We show the S/N histograms of the true noise maps (green curves) and the signal maps (gray shading) for the COSMOS field in Figure \ref{histo}. The black curves are the expected pure noise distributions with $\sigma=1$, and they agree nicely with the results of our true noise maps. The positive long tails, as well as  excess signals relative to pure noise, are from real astronomical sources. Because of the negative trough of the matched-filter PSF, 
we also see a negative tail in the distribution (\citealt{Chapin:2013fk}). 

\begin{figure}
 \begin{center}
    \leavevmode
      \includegraphics[scale=0.48]{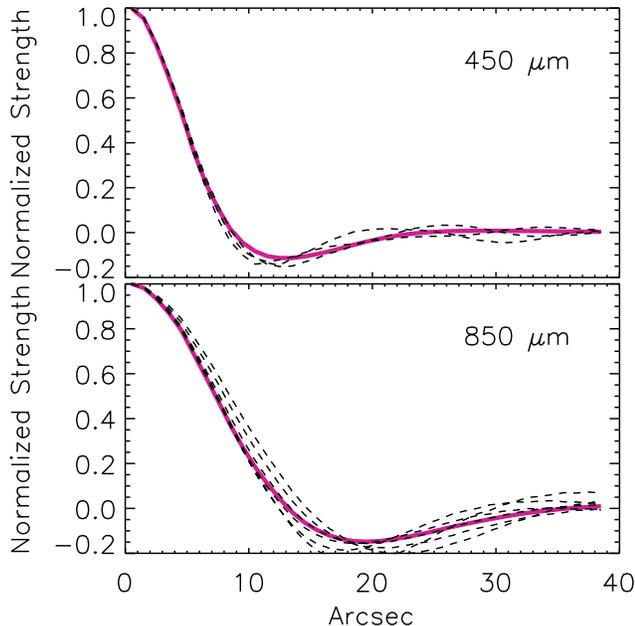}
       \caption{Normalized PSFs as a function of the distance, expressed in arcseconds relative to the center. Averaged PSFs for the A1689 maps are shown in red, and PSFs for the individual strong sources (S/N $>$ 10 at 850\,$\mu$m and S/N $>$ 7 at 450\,$\mu$m) detected in the A1689 maps are shown with dashed curves.}
     \label{psf}
  \end{center}
\end{figure}

\subsection{Methodology}

In previous work, we extracted sources down to $\sim$ 4\,$\sigma$ and used these robust catalogs with low contamination rates ($\le$ 5\%; e.g., \citealt{Chen:2013fk, Casey:2013uq}) to calculate the number counts. However, as shown in Figure \ref{histo}, excess positive signal can be seen to $\sim$ 2\,$\sigma$. For studies on individual sources, low contamination rates are essential to justify the robustness of the sample. However, for number counts analyses where positional information is no longer important, we can lower our detection threshold to a level where there are still excess counts that are statistically significant. We therefore adjusted our S/N thresholds to $\sim$2\,$\sigma$. Because the exact thresholds vary for each field, we experimented with different binning and S/N cutoffs in order to exploit fully the statistically significant signals to obtain the deepest counts possible. 

\begin{figure}
 \begin{center}
    \leavevmode
      \includegraphics[scale=0.83]{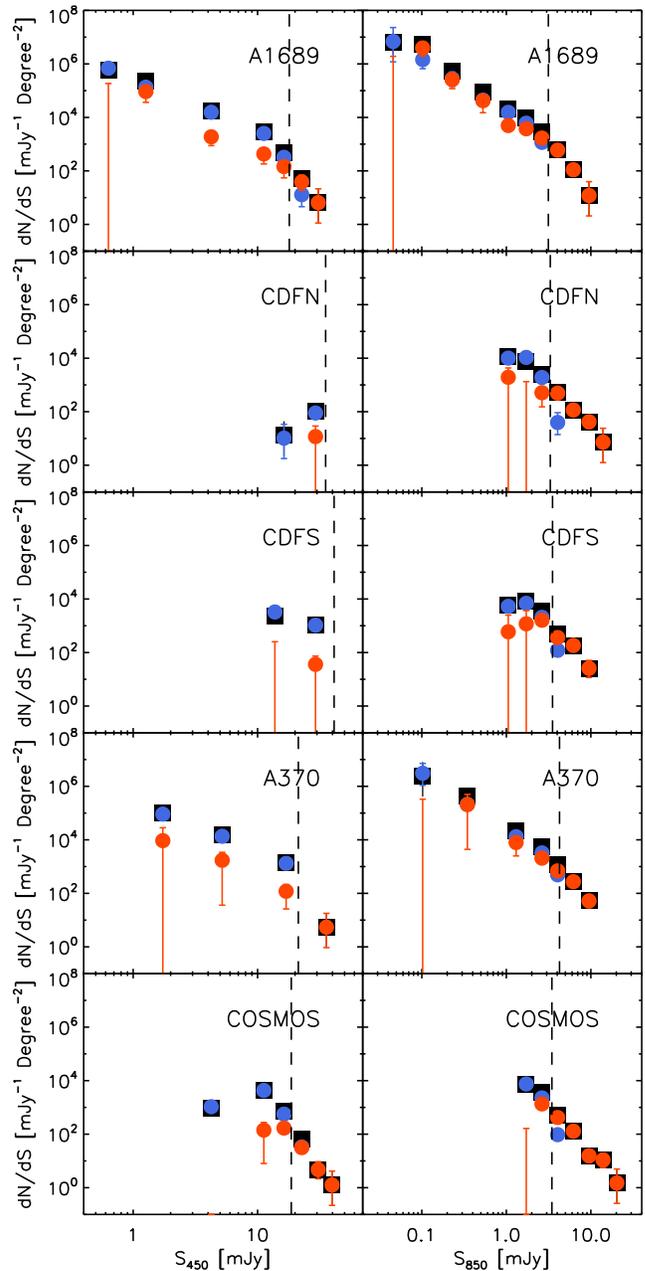}
       \caption{Differential number counts for all five fields at 450\,$\mu$m ({\it left}) and 850\,$\mu$m ({\it right}). The black (blue) symbols are the counts from the signal (true noise) maps. The red symbols are the pure source counts. The dashed vertical lines mark the mean 4\,$\sigma$ in each field. Statistically significant pure source counts can be seen below these thresholds in all the fields.}
     \label{alldefcnts}
  \end{center}
\end{figure}

Following \citet{Chen:2013fk}, we generated our source catalogs by identifying the peak S/N pixel, subtracting this peak pixel and its surrounding areas using the PSF scaled and centered on the value and position of that pixel, and then searching for the next S/N peak. We iterated this process until we hit the S/N threshold. 

We generated the PSFs used for creating the catalogs by making a weighted average of all 
primary calibrators taken before and after the science data. These are mostly Uranus, CRL618, and CRL2688.
As an example, in Figure \ref{psf} we show how
the normalized averaged PSFs of A1689 agree well with the
PSFs of the individual strong sources detected in A1689. 
The FWHM of the PSF is slightly larger than the ideal size, which could mean the observations 
were slightly out of focus. 

We ran the extraction on both the signal maps and the true noise maps. We computed the number density for each extracted source by inverting the detectable area, which is the area over which the source can be detected above the S/N threshold given the noise level. We then calculated the number counts by summing up the number densities of the sources selected in each flux bin. Finally, we subtracted the counts obtained from the true noise maps, if any, from the counts obtained from the signal maps, to produce the pure source number counts. 

\begin{figure}
 \begin{center}
    \leavevmode
      \includegraphics[scale=0.35]{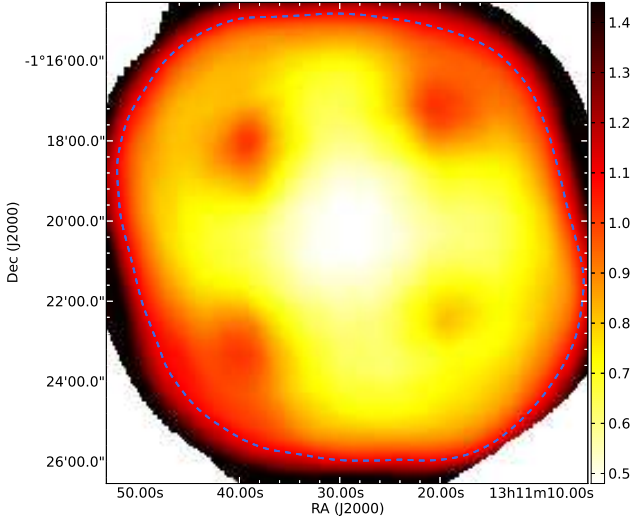}
       \caption{The r.m.s. value map for A1689 at 850\,$\mu$m in units of mJy/beam. The blue dashed contour shows 2 times the central noise.}
     \label{a1689rmsmap}
  \end{center}
\end{figure}

\begin{figure}
 \begin{center}
    \leavevmode
      \includegraphics[scale=0.68]{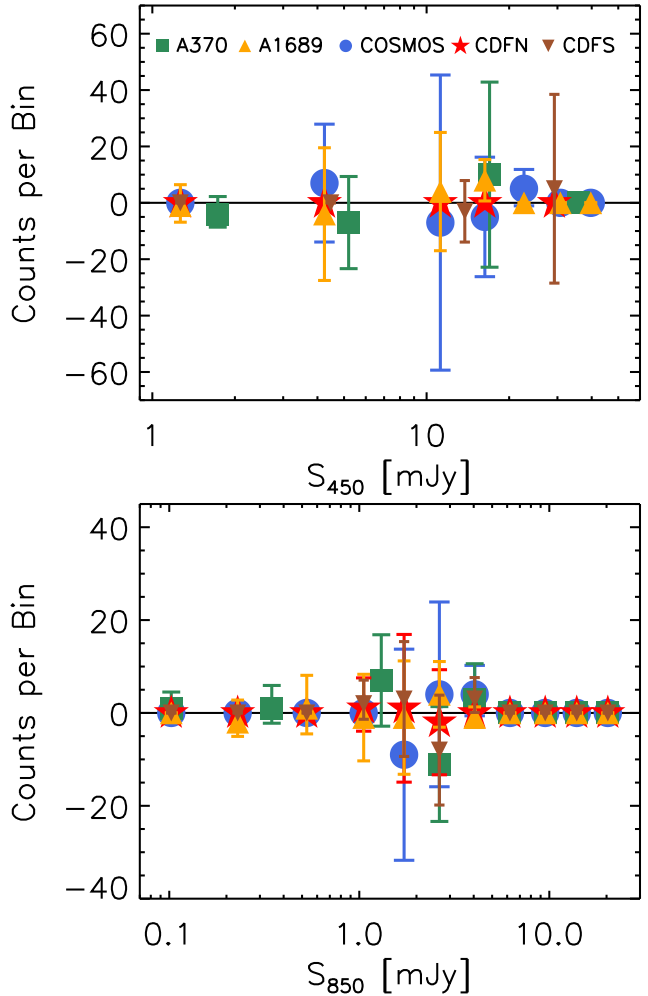}
       \caption{Differenced differential counts per bin (i.e., average counts from the simulations minus the counts from the true noise maps) on A1689 ({\it orange triangles}), CDF-N ({\it red stars}), CDF-S ({\it brown upside down triangles}), A370 ({\it green squares}), and COSMOS ({\it blue circles}) with 1\,$\sigma$ error bars. There is no statistical difference between these two sets of counts.}
     \label{nscntsdif}
  \end{center}
\end{figure}

We plot the differential number counts of all five fields in Figure \ref{alldefcnts}. The black (blue) symbols are the counts from the signal (true noise) maps, and the red symbols are the pure source counts. The vertical dashed lines represent the mean 4\,$\sigma$ depth of each field. The counts are dominated by noise at the faint end and by real sources at the bright end. The last few bins start to show signs of incompleteness. We de-lensed all the counts in the cluster fields using LENSTOOL \citep{Kneib:1996p3751} and the latest lensing models from \citet{Richard:2010fk} (A370) and \citet{Limousin:2007fj} (A1689). Thanks to the gravitational lensing in the cluster fields, we can see statistically significant pure source counts much deeper than can be probed with the 4\,$\sigma$ limits. Our statistical analysis also allows us to detect deeper counts in the blank fields.

We note that fair estimations of the pure noise counts are critical to compute the legitimate pure source counts. In SCUBA-2 maps, the noise in neighboring pixels is correlated due to the fact that as the bolometer array scans through the sky, the pixels covered by the same bolometer record the noise pattern from that bolometer. And since the noise patterns are correlated in the time domain, the noise of the pixels close to one another are correlated. Also, the performance of each subarray is different, which leads to some parts of the map being nosier than others \citep{Holland:2013lr}, and this property of the noise is retained in the true noise maps. Thus, the pure noise counts obtained from the true noise maps should be the most representative noise counts. 

However, in non-uniform maps like ours, pure noise counts based on the positions of single extractions from the true noise maps could be biased, given that occasionally some signals would happen to be located in small noisier regions, or in highly amplified regions in lensing fields. We tested for possible biased noise counts by iterating the following process : { We took the source catalogs generated 
from the true noise maps and randomized the positions of each source. In the catalogs we have the source information on S/N, fluxes, r.m.s. values and positions. 
In the process of randomization, we retained the S/N information. With the new assigned position, each source has a new r.m.s. value obtained by matching
the new position to the original r.m.s. value map. In Figure \ref{a1689rmsmap}, we show an example r.m.s. value map for A1689 at 850\,$\mu$m. The r.m.s. value maps were 
generated by computing the variance of the data that lands in each pixel. Thus the r.m.s. value maps naturally keep the information of the performance of the bolometers, 
as well as the correlations among the nearby pixels. Note also that we prevented two random positions from coming closer than half of the beam FWHM, which is the case in the real catalogs, since we removed the detected signals by subtracting a PSF scaled to the peak of the detection during our process of signal extraction. Once the new r.m.s. values were assigned, the fluxes were calculated based on the new r.m.s. values and the retained S/N ratios.
We then computed the noise counts according to the new fluxes and positions of all the sources.}

We iterated this process 50 times and calculated the average noise counts for each flux bin. We subtracted the average counts from the simulations from the counts from the true noise maps and plotted them in Figure \ref{nscntsdif}. The noise counts obtained from these two methods agree with each other to within the uncertainties for all five fields. Thus, we conclude that the pure noise counts obtained from the true noise maps are robust.

\begin{table}
\begin{center}
\caption{The resulting parameters of the broken power law model curves}
\begin{tabular}{cccrrrr}
\hline
\hline
\multirow{2}{*}{Field} & Wavelengths   &  N$_0$  &  S$_0$  &  \multirow{2}{*}{$\alpha$}  &  \multirow{2}{*}{$\beta$}  \\
  				& { ($\mu$m)}   &  { (mJy$^{-1}$ deg$^{-2}$)}  &  { ($\mu$m)}  &    &    \\
\hline
 \multirow{2}{*}{A1689}   &     450  &    30  &  20.9  &      2.5  &     6.0  \\
      &   850  &    160  &   5.45  &      2.25	  &     5.0  \\
\hline
CDF-N        & 850  &    340  &   4.5  &      1.4	  &     3.5  \\
\hline
CDF-S        & 850  &    270  &   4.5  &      2.1	  &     3.0 \\
\hline
\multirow{2}{*}{A370}        &  450  &    30  &  20.9  &      1.9  &     4.0  \\
     &    850  &    200  &   5.45  &      2.15	  &     3.0  \\
\hline
\multirow{2}{*}{COSMOS}       &   450  &    25  &  20.9  &      2.5  &     6.0  \\
    &     850  &    160  &   5.45  &      2.25	  &     3.5  \\
         
\hline
\end{tabular}
\label{bdp}
\end{center}
\end{table}

\begin{table*}
\caption{Corrected Differential Number Counts on each Individual Field}
\begin{center}
\begin{tabular}{clcl c cl c cl}
\hline
\hline
 \multicolumn{4}{c}{A1689} & &\multicolumn{2}{c}{CDF-N} & & \multicolumn{2}{c}{CDF-S} \\
 \cline{1-4}\cline{6-7}\cline{9-10}
 S$_{450}$ & dN/dS &  S$_{850}$ & dN/dS & &S$_{850}$ & dN/dS &&  S$_{850}$ & dN/dS   \\
 (mJy) & (mJy$^{-1}$ deg$^{-2}$) & (mJy)  &  (mJy$^{-1}$ deg$^{-2}$) && (mJy) & (mJy$^{-1}$ deg$^{-2}$) && (mJy)  &  (mJy$^{-1}$ deg$^{-2}$) \\

 \hline
   1.26  &  $99640\begin{array}{l}\scriptstyle +67202\\\scriptstyle -60525\end{array}$  &    0.10  &  $5436690\begin{array}{l}\scriptstyle +4016800\\\scriptstyle -2727500\end{array}$  &     &    2.64  &  $609.0\begin{array}{l}\scriptstyle +428.4\\\scriptstyle -428.4\end{array}$  &     &    2.64  &  $813.4\begin{array}{l}\scriptstyle +212.2\\\scriptstyle -212.2\end{array}$  \\
   4.24  &  $884.4\begin{array}{l}\scriptstyle +469.3\\\scriptstyle -469.3\end{array}$  &    0.23  &  $133324\begin{array}{l}\scriptstyle +103520\\\scriptstyle -72100\end{array}$      &     &    4.05  &  $410.2\begin{array}{l}\scriptstyle +95.61\\\scriptstyle -87.84\end{array}$  &     &    4.05  &  $270.2\begin{array}{l}\scriptstyle +91.11\\\scriptstyle -81.75\end{array}$  \\
  11.23  &  $171.7\begin{array}{l}\scriptstyle +99.21\\\scriptstyle -99.21\end{array}$  &    0.53  &  $36446\begin{array}{l}\scriptstyle +26232\\\scriptstyle -23530\end{array}$        &     &    6.20  &  $96.39\begin{array}{l}\scriptstyle +44.02\\\scriptstyle -31.52\end{array}$  &     &    6.20  &  $129.0\begin{array}{l}\scriptstyle +44.52\\\scriptstyle -34.11\end{array}$  \\
  16.31  &  $48.85\begin{array}{l}\scriptstyle +30.42\\\scriptstyle -30.42\end{array}$  &    1.05  &  $4729\begin{array}{l}\scriptstyle +1698\\\scriptstyle -1698\end{array}$           &     &    9.51  &  $29.04\begin{array}{l}\scriptstyle +19.64\\\scriptstyle -12.55\end{array}$  &     &    9.51  &  $23.03\begin{array}{l}\scriptstyle +22.40\\\scriptstyle -12.54\end{array}$  \\
  22.65  &  $27.01\begin{array}{l}\scriptstyle +21.38\\\scriptstyle -13.75\end{array}$  &    1.72  &  $2649\begin{array}{l}\scriptstyle +745.5\\\scriptstyle -745.5\end{array}$         &     &   13.94  &  $5.21\begin{array}{l}\scriptstyle +11.97\\\scriptstyle -4.31\end{array}$    &     &  \ldots  &  \ldots                                                                      \\
  30.74  &  $5.41\begin{array}{l}\scriptstyle +12.43\\\scriptstyle -4.47\end{array}$    &    2.64  &  $886.0\begin{array}{l}\scriptstyle +219.1\\\scriptstyle -219.1\end{array}$        &     &  \ldots  &  \ldots                                                                      &     &  \ldots  &  \ldots                                                                      \\
 \ldots  &  \ldots                                                                      &    4.05  &  $313.4\begin{array}{l}\scriptstyle +68.39\\\scriptstyle -68.39\end{array}$        &     &  \ldots  &  \ldots                                                                      &     &  \ldots  &  \ldots                                                                      \\
 \ldots  &  \ldots                                                                      &    6.20  &  $127.7\begin{array}{l}\scriptstyle +76.30\\\scriptstyle -50.67\end{array}$        &     &  \ldots  &  \ldots                                                                      &     &  \ldots  &  \ldots                                                                      \\
 \ldots  &  \ldots                                                                      &    9.51  &  $8.02\begin{array}{l}\scriptstyle +18.45\\\scriptstyle -6.63\end{array}$          &     &  \ldots  &  \ldots                                                                      &     &  \ldots  &  \ldots                                                                      \\
\hline
\hline
\multicolumn{4}{c}{A370} && \multicolumn{5}{c}{COSMOS} \\
 \cline{1-4}\cline{6-10}
S$_{450}$ & dN/dS &  S$_{850}$ & dN/dS & &S$_{450}$ & dN/dS & &  S$_{850}$ & dN/dS \\
 (mJy) & (mJy$^{-1}$ deg$^{-2}$) & (mJy)  &  (mJy$^{-1}$ deg$^{-2}$) && (mJy) & (mJy$^{-1}$ deg$^{-2}$) && (mJy)  &  (mJy$^{-1}$ deg$^{-2}$) \\
\hline
   5.20  &  $573.1\begin{array}{l}\scriptstyle +561.1\\\scriptstyle -561.1\end{array}$  &    0.35  &  $145886\begin{array}{l}\scriptstyle +203320\\\scriptstyle -142801\end{array}$     &     &   11.23  &  $55.85\begin{array}{l}\scriptstyle +52.68\\\scriptstyle -52.68\end{array}$  &     &    2.64  &  $705.1\begin{array}{l}\scriptstyle +134.0\\\scriptstyle -134.0\end{array}$  \\
  16.97  &  $106.7\begin{array}{l}\scriptstyle +83.48\\\scriptstyle -83.48\end{array}$  &    1.31  &  $8384\begin{array}{l}\scriptstyle +5740\\\scriptstyle -5740\end{array}$           &     &   16.31  &  $55.01\begin{array}{l}\scriptstyle +18.56\\\scriptstyle -18.56\end{array}$  &     &    4.05  &  $248.4\begin{array}{l}\scriptstyle +41.82\\\scriptstyle -39.62\end{array}$  \\
  35.78  &  $2.03\begin{array}{l}\scriptstyle +4.67\\\scriptstyle -1.68\end{array}$     &    2.64  &  $1485\begin{array}{l}\scriptstyle +434.4\\\scriptstyle -434.4\end{array}$         &     &   22.65  &  $11.55\begin{array}{l}\scriptstyle +4.55\\\scriptstyle -4.55\end{array}$    &     &    6.20  &  $118.4\begin{array}{l}\scriptstyle +22.78\\\scriptstyle -22.78\end{array}$  \\
 \ldots  &  \ldots                                                                      &    4.05  &  $386.7\begin{array}{l}\scriptstyle +122.6\\\scriptstyle -113.6\end{array}$        &     &   30.74  &  $2.18\begin{array}{l}\scriptstyle +2.12\\\scriptstyle -1.19\end{array}$     &     &    9.51  &  $10.97\begin{array}{l}\scriptstyle +7.42\\\scriptstyle -4.74\end{array}$    \\
 \ldots  &  \ldots                                                                      &    6.20  &  $239.7\begin{array}{l}\scriptstyle +73.32\\\scriptstyle -57.53\end{array}$        &     &   39.69  &  $0.47\begin{array}{l}\scriptstyle +1.09\\\scriptstyle -0.39\end{array}$     &     &   13.94  &  $9.84\begin{array}{l}\scriptstyle +7.78\\\scriptstyle -4.71\end{array}$     \\
 \ldots  &  \ldots                                                                      &    9.51  &  $37.94\begin{array}{l}\scriptstyle +25.66\\\scriptstyle -16.39\end{array}$        &     &  \ldots  &  \ldots                                                                      &     &   20.31  &  $1.50\begin{array}{l}\scriptstyle +3.46\\\scriptstyle -1.24\end{array}$     \\
\hline
 \end{tabular}
\label{eachcnts}
\end{center}
\end{table*}

Following \citet{Chen:2013fk}, we ran Monte Carlo simulations to estimate the underlying counts models. 
We first randomly populated the true noise map with simulated sources, drawn from an assumed model 
and convolved with the PSFs, to form a simulated image. 
The counts model is in the form of a broken power law 
\begin{equation}
  \frac{dN}{dS} = \left\{
  \begin{array}{l l}
    {N_0}\left(\frac{S}{S_0}\right)^{-\alpha}  & \quad \text{if $S \leq S_0$}\\
    {N_0}\left(\frac{S}{S_0}\right)^{-\beta}  & \quad \text{if $S > S_0$} \\
  \end{array} \right.
\end{equation}

The faintest fluxes we adopted for any of our models are the fluxes at which the integrated flux density agrees with the EBL measurements within errors \citep{Puget:1996p2082,Fixsen:1998p2076}. We then extracted the signal and computed the recovered number counts in exactly the same way as we did with the real data maps.  We measured the ratio between the recovered counts and the input counts, which reflects the Eddington bias (\citealt{Eddington:1913fj}), and then applied this ratio to the statistically significant observed counts to correct for that bias. We did a $\chi^2$ fit to the corrected observed counts using a broken power law to obtain the normalization and power law indices. We used this fit as the next iteration of the model counts in
the procedure and repeated the process. We continued until
the input model agreed with the corrected counts at the
1\,$\sigma$ level throughout the statistically significant
range.

There are only three (four) statistically significant points in A370 450\,$\mu$m (CDF-S 850\,$\mu$m) counts, so we only fitted the normalization to avoid overfitting. {We also note that the fitting results were not affected by specifying the break positions of the broken power laws into the fit. We chose to fix the breaks to again avoid overfitting and to obtain better statistical constraints on the fit.}
We excluded the CDF-N and CDF-S from the 450\,$\mu$m analysis, because we have no statistically significant 450\,$\mu$m counts for these fields. 

\begin{figure}
 \begin{center}
    \leavevmode
      \includegraphics[scale=0.68]{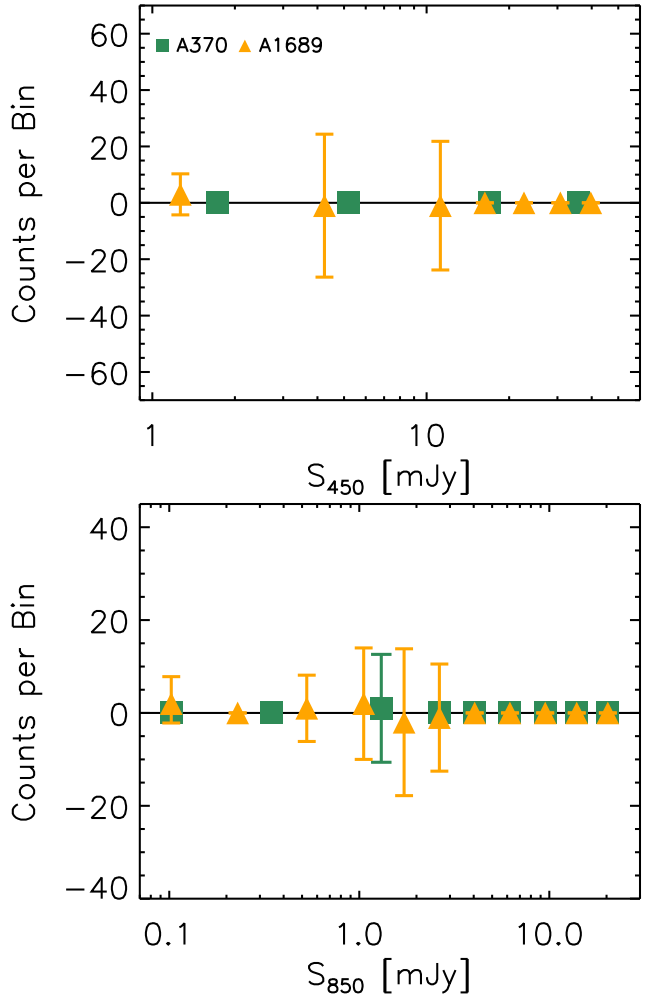}
       \caption{Differenced differential counts per bin (i.e., average counts from the simulations minus the counts from the real signal maps) for A1689 ({\it orange triangles}) and A370 ({\it green squares}) with 1\,$\sigma$ error bars. There is no statistical difference between these two sets of counts.}
     \label{clustecntsdif}
  \end{center}
\end{figure}

\begin{figure}
 \begin{center}
    \leavevmode
      \includegraphics[scale=0.75]{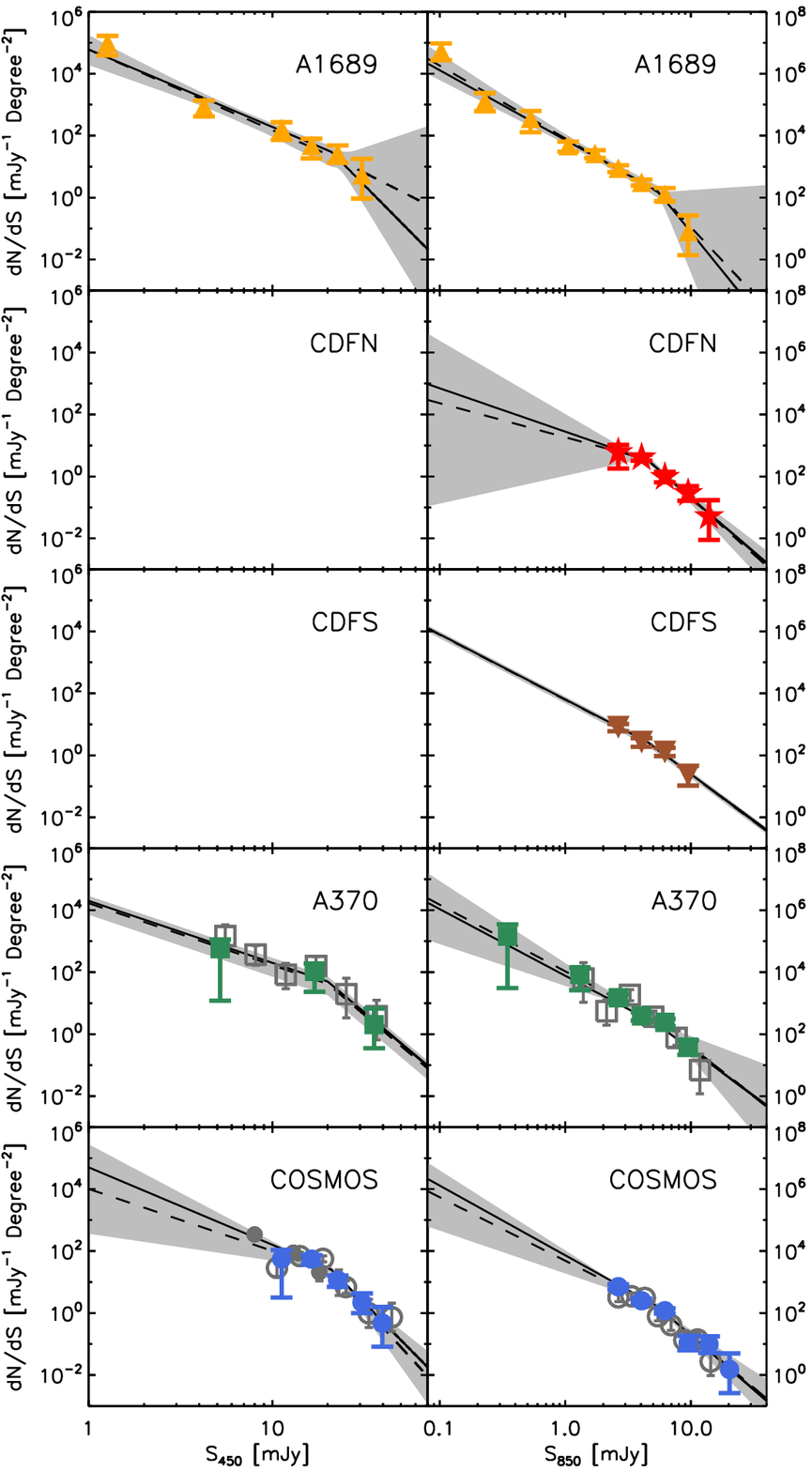}
       \caption{Corrected differential number counts (colored symbols) for all five fields. The black curves are the input counts models in our Monte Carlo simulations. The dashed curves show the minimum $\chi^2$ fits. The 1\,$\sigma$ error regions are illustrated by gray shading. The empty gray squares are SCUBA-2 counts from \citet{Chen:2013fk}, and the empty (filled) gray circles are SCUBA-2 counts from \citet{Casey:2013uq} \citep{Geach:2013kx}. There are no statistically significant counts detected at 450\,$\mu$m for CDF-N and CDF-S.
}
     \label{simall}
  \end{center}
\end{figure}

For the cluster fields, we populated the simulated sources in the source plane and imaged them onto the image plane using LENSTOOL. 
At 850\,$\mu$m, we located the source planes at $z=3.0$ based on the latest observational results \citep{Barger:2012lr, Vieira:2013vn} and theoretical models \citep{Hayward:2013lr}. 
At 450\,$\mu$m, two recent SCUBA-2 results on the COSMOS field have shown that the majority of the 450\,$\mu$m sources are at $z < 3$ \citep{Geach:2013kx, Casey:2013uq}. 
We located the 450\,$\mu$m source planes at $z=1.3$ based on \citet{Geach:2013kx}, since their flux range is closer that of our observations. 
In any case, we stress that our statistical approach to estimating the counts is not sensitive to the adopted source plane redshifts \citep{Blain:1999p7279}.  

{On the other hand, however, the de-lensing process could be significantly affected by the uncertainties of the source positions \citep{Chen:2011p11605}. We tested this bias by again running the Monte Carlo simulations on the extracted signals by randomizing their positions. We showed in \citet{Chen:2013fk} that the positional uncertainty is a function of S/N; thus, we randomized the position of each source by setting the offsets smaller than the 90\% confidence boundary given its S/N. We then calculated the de-lensed counts based on the new positions. We iterated this process 50 times and obtained the average source counts. We then subtracted the average counts from the counts obtained from the real signal maps. We show our results in Figure \ref{clustecntsdif}. The counts obtained from these two methods agree with each other to within the uncertainties for both A1689 and A370. This illustrates the robustness  of our methodology. Given the small effects of the positional uncertainties on the de-lensing process, we conclude that using our statistical method, the uncertainty on the overall counts caused by the de-lensing process is negligible. }

\begin{figure}
 \begin{center}
    \leavevmode
      \includegraphics[scale=0.69]{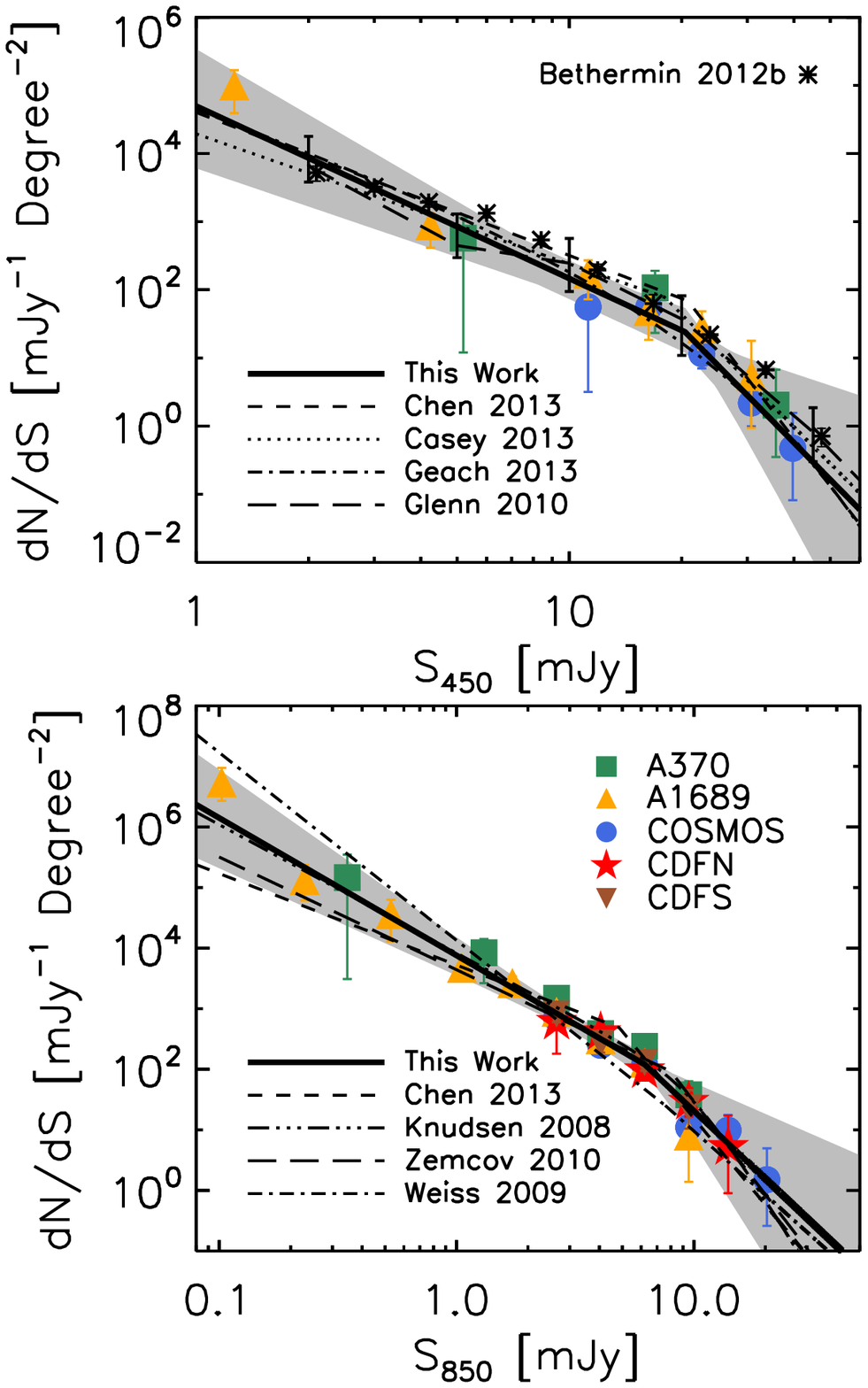}
       \caption{Differential number counts for all five fields (colored symbols) at 450\,$\mu$m ({\it upper}) and 850\,$\mu$m ({\it lower}). Solid black curves are best $\chi^2$ broken power law fits with error regions in gray shading (Table \ref{fr}). Black dashed curves represent the SCUBA-2 counts from \citet{Chen:2013fk}. At 450\,$\mu$m, the counts models from several other works are plotted as dotted \citep{Casey:2013uq} and dot-dashed \citep{Geach:2013kx} curves. The multiply-broken power law model for the 500\,$\mu$m counts computed through a P(D) analysis on {\it Herschel} maps is shown as the long-dashed curve with error bars \citep{Glenn:2010kx}, and the asterisks are the hybrid {\it Herschel} 500\,$\mu$m counts \citep{Bethermin:2012yq}. At 850\,$\mu$m, two counts models obtained from SCUBA cluster fields surveys are shown as the dot-dot-dot-dashed \citep{Knudsen:2008p3824} and long-dashed \citep{Zemcov:2010uq} curves in the lower panel. The counts model from the LABOCA survey on the ECDF-S \citep{Weis:2009qy} is plotted as the dot-dashed curve. }
     \label{allcnts}
  \end{center}
\end{figure}

\subsection{Results}

\begin{table}
\begin{center}
\caption{Best $\chi^2$ fits on the number counts from all five fields at 450 and 850 micron}
\begin{tabular}{cccccc}
\hline
\hline
 Wavelengths   &  N$_0$  &  S$_0$  &  \multirow{2}{*}{$\alpha$}  &  \multirow{2}{*}{$\beta$}  \\
($\mu$m)   &  (mJy$^{-1}$ deg$^{-2}$)  & (mJy)  &    &    \\
\midrule
\vspace{1mm}\
         450  &    $24^{+30}_{-12}$  &  20.4  &      $2.53^{+0.67}_{-0.67}$ &    $ 5.57^{+4.5}_{-4.0}$  \\
\vspace{1mm}\
         850  &    $120^{+65}_{-45}$  &   6.21  &      $2.27^{+0.5}_{-0.5}$ &     $3.71^{+2.5}_{-2.0}$ \\
\hline
\end{tabular}
\label{fr}
\end{center}
\end{table}

We show the corrected differential number counts for each field individually in Figure \ref{simall} (colored symbols). This figure clearly illustrates the advantage of observing both blank and cluster fields in order to probe a wide flux range. The black solid curves show the input models. The dashed curves show the minimum $\chi^2$ fits. The gray shading denotes the 1\,$\sigma$ error regions based on the fits. We also show the SCUBA-2 counts constructed using a 4\,$\sigma$ detection threshold
for A370 \citep{Chen:2013fk} and a 3.8\,$\sigma$ detection threshold for COSMOS \citep{Casey:2013uq, Geach:2013kx} (gray symbols), in which they both extrapolated their counts to fainter end by doing source flux deboosting. These agree nicely with our results. In Tables \ref{bdp} and \ref{eachcnts}, we summarize, respectively, the model parameters and the corrected counts for each field .

\begin{table}
\begin{center}
\caption{Best $\chi^2$ fits on the combined differential number counts at 450 and 850 micron}
\begin{tabular}{cccccc}
\hline
\hline
 Wavelengths   &  N$_0$  &  S$_0$  &  \multirow{2}{*}{$\alpha$}  &  \multirow{2}{*}{$\beta$}  \\
($\mu$m)   &  (mJy$^{-1}$ deg$^{-2}$)  & (mJy)  &    &    \\
\midrule
\vspace{1mm}\
         450  &    $22^{+51}_{-15}$  &  $21.1^{+9.0}_{-7.0}$&      $2.73^{+0.0}_{-0.0}$ &    $ 5.77^{+9.5}_{-2.00}$  \\
\vspace{1mm}\
         850  &    $120^{+180}_{-70}$  &   $6.2^{+2.0}_{-2.0}$  &      $2.26^{+0.2}_{-0.2}$ &     $3.79^{+2.00}_{-1.00}$ \\
\hline
\end{tabular}
\label{cmcntmdl}
\end{center}
\end{table}

\begin{table}
\begin{center}
\caption{The combined differential number counts at 450 and 850 micron}
\begin{tabular}{rlrl}
\hline
\hline
 S$_{450}$  &    dN/dS                                                                       &  S$_{850}$  &  dN/dS                                                                   \\
      (mJy)  &  (mJy$^{-1}$ deg$^{-2}$)                                                     &        (mJy)  &  (mJy$^{-1}$ deg$^{-2}$)                                                                 \\
\hline
   1.26  &  $99640\begin{array}{l}\scriptstyle +67202\\\scriptstyle -60525\end{array}$  &   0.10  &  $5436690\begin{array}{l}\scriptstyle +4016800\\\scriptstyle -2727500\end{array}$  \\
   4.24  &  $1504\begin{array}{l}\scriptstyle +689.2\\\scriptstyle -689.2\end{array}$   &   0.23  &  $133324\begin{array}{l}\scriptstyle +103520\\\scriptstyle -72100\end{array}$      \\
  11.23  &  $78.74\begin{array}{l}\scriptstyle +47.92\\\scriptstyle -47.92\end{array}$  &   0.53  &  $36446\begin{array}{l}\scriptstyle +26232\\\scriptstyle -23530\end{array}$        \\
  16.31  &  $58.09\begin{array}{l}\scriptstyle +16.95\\\scriptstyle -16.95\end{array}$  &   1.05  &  $4729\begin{array}{l}\scriptstyle +1698\\\scriptstyle -1698\end{array}$           \\
  22.65  &  $14.60\begin{array}{l}\scriptstyle +4.77\\\scriptstyle -4.77\end{array}$    &   1.72  &  $2021\begin{array}{l}\scriptstyle +553.1\\\scriptstyle -553.1\end{array}$         \\
  30.74  &  $2.82\begin{array}{l}\scriptstyle +2.23\\\scriptstyle -1.35\end{array}$     &   2.64  &  $829.4\begin{array}{l}\scriptstyle +101.3\\\scriptstyle -101.3\end{array}$        \\
  39.69  &  $0.47\begin{array}{l}\scriptstyle +1.09\\\scriptstyle -0.39\end{array}$     &   4.05  &  $308.8\begin{array}{l}\scriptstyle +31.50\\\scriptstyle -31.50\end{array}$        \\
 \ldots  &  \ldots                                                                      &   6.20  &  $131.1\begin{array}{l}\scriptstyle +15.34\\\scriptstyle -15.34\end{array}$        \\
 \ldots  &  \ldots                                                                      &   9.51  &  $20.03\begin{array}{l}\scriptstyle +5.73\\\scriptstyle -4.55\end{array}$          \\
 \ldots  &  \ldots                                                                      &  13.94  &  $8.62\begin{array}{l}\scriptstyle +5.83\\\scriptstyle -3.72\end{array}$           \\
 \ldots  &  \ldots                                                                      &  20.31  &  $1.50\begin{array}{l}\scriptstyle +3.46\\\scriptstyle -1.24\end{array}$           \\
\hline
\end{tabular}
\label{cmcnt}
\end{center}
\end{table}

In Figure \ref{allcnts}, we show all the counts together for the two wavelengths. The black solid curves represent the best fit models, which we present in Table \ref{fr}. At 850\,$\mu$m, our results are almost indistinguishable from the SCUBA results in \citet{Knudsen:2008p3824} ({\it dot-dot-dot-dashed curve}) covering a similar flux range, and they agree within the errors with the \citet{Zemcov:2010uq} results {\it (long-dashed)}, which come from an analysis of all the SCUBA data taken on cluster fields. We do not observe a significant under-abundance in the CDF-S 850\,$\mu$m counts above 3\,mJy, as was found in the LABOCA wider area but shallower sensitivity survey of the Extended CDF-S (ECDF-S) \citep{Weis:2009qy} {\it (dot-dashed)}. We discuss potential causes for this discrepancy in Section~\ref{subsec:cdfs}. The SCUBA-2 counts for A370 (\citealt{Chen:2013fk}; {\it short-dashed}) are slightly higher relative to the other fields at $\sim3-6$\,mJy (850\,$\mu$m) and $\sim10-25$\,mJy (450\,$\mu$m), which could be caused by { sample variance from fields which are small in size compared to the large-scale structure}. We show the hybrid {\it Herschel} 500\,$\mu$m counts \citep{Bethermin:2012yq} obtained from a mix of directly resolved counts above 20\,mJy and 24\,$\mu$m stacking counts below 20\,mJy with asterisks. The noticeable over-abundance of those counts relative to ours could indicate that the {\it Herschel} counts are biased upward due to source blending caused by poorer resolution ($\sim$35$''$ beam FWHM compared to 7$\farcs$5 beam FWHM). We also show various other results at 450\,$\mu$m from the literature, and they all agree well with our results.

\begin{figure}
 \begin{center}
    \leavevmode
      \includegraphics[scale=0.35]{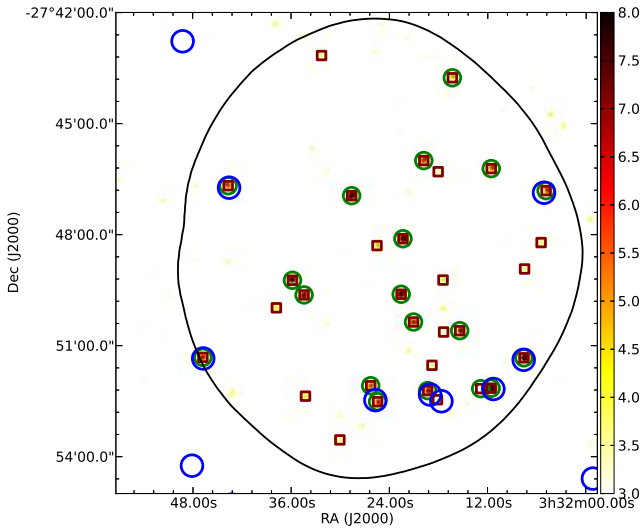}
       \caption{SCUBA-2 signal-to-noise map of CDF-S at 850\,$\mu$m with {\it blue circles} marking the LABOCA detections \citep{Weis:2009qy} and {\it brown squares} showing our 4\,$\sigma$ detections. {\it Green circles} represent the sources that are significantly detected ($>\,4\,\sigma$) by SCUBA-2 and also should be detectable by the LABOCA observations given their fluxes. The black curve encloses the effective area we adopted for the number counts calculation, which is 2 times the central sensitivity.}
     \label{cdfs}
  \end{center}
\end{figure}

\section{Possible issues}
\subsection{Field-to-field Variance: Is the CDF-S Underdense?}
\label{subsec:cdfs}

\begin{figure}
 \begin{center}
    \leavevmode
      \includegraphics[scale=0.55]{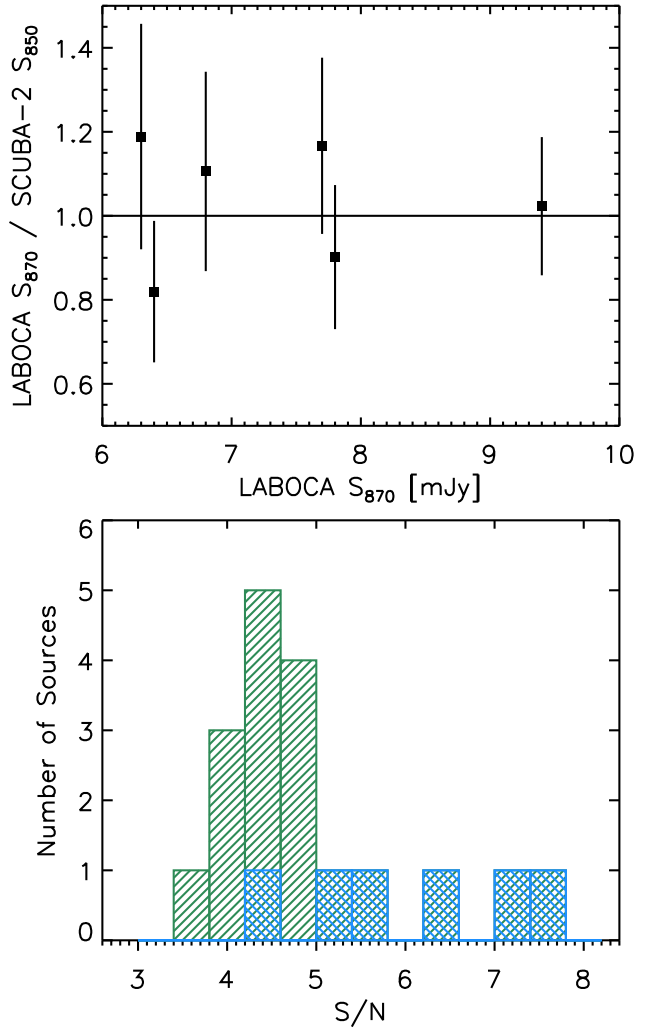}
       \caption{{\it Upper}: The ratio of LABOCA 870\,$\mu$m to SCUBA-2 850\,$\mu$m fluxes on the sources that are detected in both observations. {\it Lower}: The {\it green} histogram shows the expected S/N for the SCUBA-2 detected sources with LABOCA detectable fluxes, based on the SCUBA-2 fluxes and the LABOCA noise claim. The {\it blue} histogram shows the same information for the fraction that are LABOCA detected.}
     \label{stonhisto}
  \end{center}
\end{figure}

Several studies of the CDF-S have shown that at redshifts between 2 and 3, the massive red galaxies selected through rest-frame optical colors (DRGs and pBzKs) that constitute the  bulk of the mass during that epoch are under-abundant relative to the mean density from other deep fields \citep{van-Dokkum:2006lr,Marchesini:2007fk,Blanc:2008qy}. However, in the same studies, there is no sign of an underdensity of non-DRGs \citep{Marchesini:2007fk} and less massive star-forming BzKs (sBzKs; \citealt{Blanc:2008qy}). Together with the fact that recent 4~Ms {\it Chandra} observations also show no sign of an underdensity \citep{Lehmer:2012uq} at the lower flux end where high-redshift late-type galaxies start to dominate the X-ray number counts, this could imply that at $2 < z < 3$, massive passive galaxies are underdense, while less massive star-forming galaxies are not.  

On the other hand, an underdensity of the 870\,$\mu$m sources was reported by the LABOCA survey of the ECDF-S \citep{Weis:2009qy}. However, we do not confirm this result. As shown in Figure \ref{allcnts}, our SCUBA-2 observations at 850\,$\mu$m of the central region of the LABOCA field show no sign of an underdensity. {Recently, \citet{Scott:2010pp} also found no apparent underdensity at 1.1\,mm from an AzTEC survey toward a similar region as our coverage. Contrary to the AzTEC observations,} our observed waveband is close to that of LABOCA, so both should see a similar population. Although the coverage of the LABOCA observations is wider (30$' \times 30'$) with slightly shallower but uniform sensitivity ($\sim$1.2 mJy/beam), an underdensity in the central region of the LABOCA map should be apparent in our data. 

In Figure \ref{cdfs}, we show our 850\,$\mu$m S/N map with the LABOCA detections plotted with blue circles and our 4\,$\sigma$ detections with brown squares. While our SCUBA-2 observations recover all the LABOCA sources, many LABOCA detectable SCUBA-2 sources (sources with fluxes greater than the LABOCA limit; green circles) are missed by LABOCA. There are 19 SCUBA-2 sources with fluxes greater than 4.4\,mJy (the LABOCA limit), only 8 of which are detected by LABOCA. If we correct for this factor of 2.375 in the cumulative counts, then the LABOCA counts 
would be in good agreement with the measurements from other fields. 

To investigate the possible causes of this discrepancy, we first compared the fluxes of the sources detected in both observations. We found that both measurements agree very well, as shown in the upper panel of Figure \ref{stonhisto}. Note that we excluded two close LABOCA sources from the analyses described in this paragraph (two overlapped blue circles in Figure \ref{cdfs}), because they are blended in the LABOCA maps, and thus the uncertainties on their flux measurements are high. We then examined the histogram of the expected S/N on the SCUBA-2 detected sources with LABOCA detectable fluxes, based on the SCUBA-2 fluxes and the LABOCA noise claim (green hatched regions in Figure \ref{stonhisto}). We found that while LABOCA succeeded in detecting all of the S/N$>$5 sources (blue hatched region), LABOCA failed to detect all but one of the lower S/N sources.
Ideally, if the two observations had agreed, then the two distributions would be identical. 

{On the other hand, the extra detections in our maps are all at very high S/N ($>$ 5), and
many of them are detected in nightly maps. We cross correlated them with the sources
detected at other wavelengths. Of our 12 SCUBA-2 sources that are not also detected in the LABOCA survey, 10 have
either 24 micron or radio counterparts, and 6 are detected at both 24 micron
and radio. Moreover, 6 are detected by AzTEC at 1.1\,mm. The fact that our extra detections are highly significant
and highly correlated with sources detected at other wavebands makes it extremely unlikely that many of them are spurious.
}

\subsection{The Effects of Multiplicity}
\label{cdfnsma}

Recently, many high resolution ($\sim$1$''$ beam FWHM) submillimeter and millimeter interferometric observations have shown that a significant percentage ($20-40$\%) of single-dish detected submillimeter sources are in fact composed of two or sometimes three separated sources (e.g., \citealt{Wang:2011p9293, Smolcic:2012pp, Barger:2012lr, Hodge:2013lr}). In semi-analytical simulations it has been shown that multiplicity could dramatically impact the number counts at 850\,$\mu$m obtained from single-dish observations \citep{Hayward:2013lr}. However, the observational constraints on the fraction of multiples are subject to small number statistics and heterogeneity in the sample selection. Above an 850\,$\mu$m flux of 7\,mJy, three of the eight SMA observed SCUBA sources in the CDF-N were found to be multiples by \citet{Barger:2012lr}, which corresponds to a multiple fraction of 37.5\% with a $\pm$1\,$\sigma$ range from $17-74$\%. \citet{Smolcic:2012pp} compiled a list of millimeter and submillimeter interferometric continuum follow-up observations of a sample of 36 LABOCA sources at 870\,$\mu$m in the COSMOS field, and they reported that 6 of their 27 interferometric detections are multiples (22\%$\pm$9). With a larger sample size, \citet{Hodge:2013lr} reported that 24 of their 69 ALMA robustly detected and LABOCA pre-selected sources (MAIN ALESS sample) are multiples, giving a multiple fraction of $\sim$35$\pm$7\%. While the FWHM of the LABOCA beam is about 37\% larger than that of SCUBA-2 (19$\farcs$2 versus 14$''$), it is unclear how much this difference in spatial resolution affects the multiple fraction. If the spatial distribution among close multiples were random, as shown by \citet{Hodge:2013lr}, then the multiple fraction for the SCUBA-2 850\,$\mu$m selected sources would be $\sim$19\% based on scaling the ALMA multiplicity of the MAIN ALESS sample. 

We have obtained Submillimeter Array (SMA) observations (all are detections) of {8} new SCUBA-2 850\,$\mu$m sources in the CDF-N with arcsecond spatial resolution. Together with previous SMA observations of SCUBA detected sources in the same field (mostly from \citealt{Barger:2012lr}), we have compiled a list of  24 SCUBA-2 detected (4\,$\sigma$) and SMA observed sources in the CDF-N. A detailed analysis is given in A.~Barger et al.\ (2013, in preparation). All of our new SMA observations are single source detections, and only {3 out of 24} SMA detected sources with $S_{850} > 3.5\,$mJy break into close pairs; that is, a multiple fraction of { 12.5$^{ +12.1}_{ -6.8}$\%}. This is consistent with the lower end of the results discussed above. 

\begin{figure}
 \begin{center}
    \leavevmode
      \includegraphics[scale=0.43]{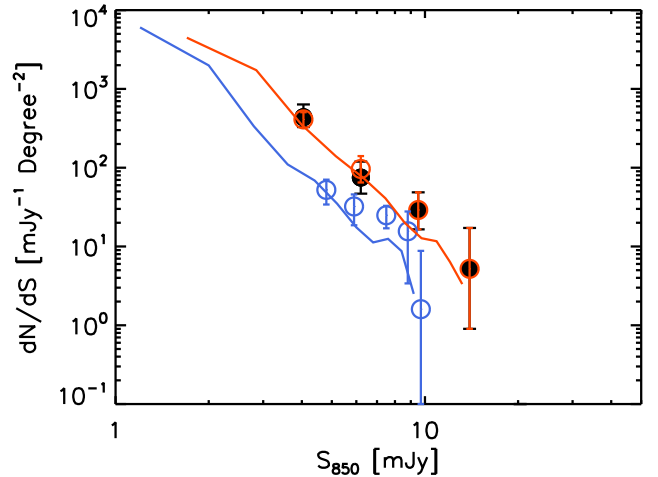}
       \caption{CDF-N 850\,$\mu$m differential number counts above 3.5\,mJy obtained from SCUBA-2 (red circles) and those inferred from the SMA observed SCUBA-2 sources (black circles). The latter are essentially corrected for the effects of multiplicity. The CDF-S 850\,$\mu$m counts obtained by \citet{Karim:2013fk} from their ALMA observations of the LABOCA survey are shown in blue. Model predictions by \citet{Hayward:2013lr} for single-dish and interferometric counts are shown with the red and blue curves, respectively.}
     \label{smacounts}
  \end{center}
\end{figure}

We recomputed the multiplicity-corrected CDF-N 850\,$\mu$m number counts above 3.5\,mJy based on these observations using
 \begin{equation}
 \begin{split}
  \frac{dN_{corr,i}(S)}{dS} = \frac{dN_{orig,i}(S)}{dS} \times (1-f_{mul,i}(S))\\
  +f_{mul,i}(2S)\times4\times\frac{dN_{orig,i}(2S)}{dS},
\end{split}
\end{equation}
where $f_{mul,i} = N_{mul}/N_{sma}$ represents the multiple fraction of the SMA detected SCUBA-2 sources in each flux bin $i$, and $\frac{dN_{corr,i}}{dS}$ and $\frac{dN_{orig,i}}{dS}$ are the multiplicity-corrected and the original SCUBA-2 counts, respectively. We have assumed for simplicity that the source splits into two equal
components. This procedure conserves the EBL contribution of the counts, as it should.

We plot the results (black circles) in Figure \ref{smacounts}, along with the original CDF-N SCUBA-2 counts (red circles). The multiplicity-corrected counts are essentially unchanged in the first two bright bins, since all the SMA observations on these sources correspond to single detections. In the last two faint bins, owing to the effect of the multiplicity, the corrected counts differ by a small amount from the original ones. However, the systematic changes introduced by
the multiplicity are smaller than the statistical noise in our counts determination. 

We compare our corrected counts with the ones obtained by the ALMA follow-up observations of the LABOCA sources (blue circles in Figure \ref{smacounts}; \citealt{Karim:2013fk}). The ALMA/LABOCA counts are systematically lower than ours, especially at the bright end above 9\,mJy, where they argued that the counts drop significantly due to the fact that the all of their bright sources tend to break into fainter multiple sources. This tendency is not confirmed by our observations. In fact, all of our SMA observed single-dish detected bright sources ($>$9\,mJy) show single SMA detections. In addition, the low counts  in the ALMA/LABOCA determination are a partial consequence of the low LABOCA counts in the CDF-S, which we discussed in the previous subsection.

\begin{figure}
 \begin{center}
    \leavevmode
      \includegraphics[scale=0.44]{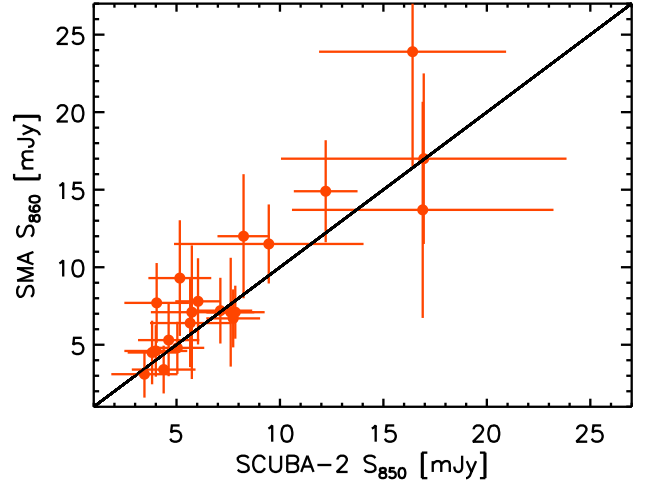}
       \caption{SMA 860\,$\mu$m flux versus SCUBA-2 850\,$\mu$m flux for the CDF-N SCUBA-2 4\,$\sigma$ detected sources with single SMA detections. Error bars show 2\,$\sigma$ uncertainties. }
     \label{flxcomp}
  \end{center}
\end{figure}

The multiple fraction could be affected by the depth of the follow-up observations. In principle, the deeper the interferometric observations, the more likely that a multiple system would be revealed. The sensitivity of our SCUBA-2 and SMA observations are comparable, with a mean depth ratio of $\sim$1.1. By comparing the fluxes measured by SCUBA-2 and by the SMA on the sources with a single SMA detection, we can estimate how many sources could be further resolved into multiples if the follow-up interferometric observations were deeper. We show the flux comparison in Figure \ref{flxcomp}, where most of the SMA measurements statistically agree with the those made by SCUBA-2. That a significant number of SMA measurements are larger than the SCUBA-2 measurements could be caused by the calibration uncertainties (The average flux ratio of the SMA to the SCUBA-2 is $\sim$1.1). 
We cannot rule out the possibility that sources with fluxes fainter 
than the sensitivity limit of our observations are
contributing a small fraction of the flux of a SCUBA-2 source and the multiple fraction could be higher, but  
such faint sources would be unlikely to affect our counts within 
our flux range of interest.

In our counts calculation, we did assume that the multiple fraction obtained from the SMA observations is applicable to the overall sample. We caution that our counts below $\sim$5\,mJy could change significantly if the multiple fraction were much higher, since our SMA observations primarily targeted the brighter SCUBA-2 sources.

We also compare our counts with the semi-analytical models recently presented by \citet{Hayward:2013lr} in Figure \ref{smacounts}. Their single-dish (they adopted a 15$''$ FWHM beam size) counts predictions are plotted as the red curve, and their predictions for counts derived with arcsecond resolution 
interferometric observations are plotted as the blue curve. 
While their single-dish predictions agree very nicely with our results, their interferometric predictions are in general agreement with the ALMA counts but are significantly lower than ours. Our results and our discussion of the \citet{Karim:2013fk} results suggest that the models proposed by \citet{Hayward:2013lr} may be overestimating the
multiplicity correction.

\section{Discussion}
\label{secdisc}

\begin{figure}
 \begin{center}
    \leavevmode
      \includegraphics[scale=0.65]{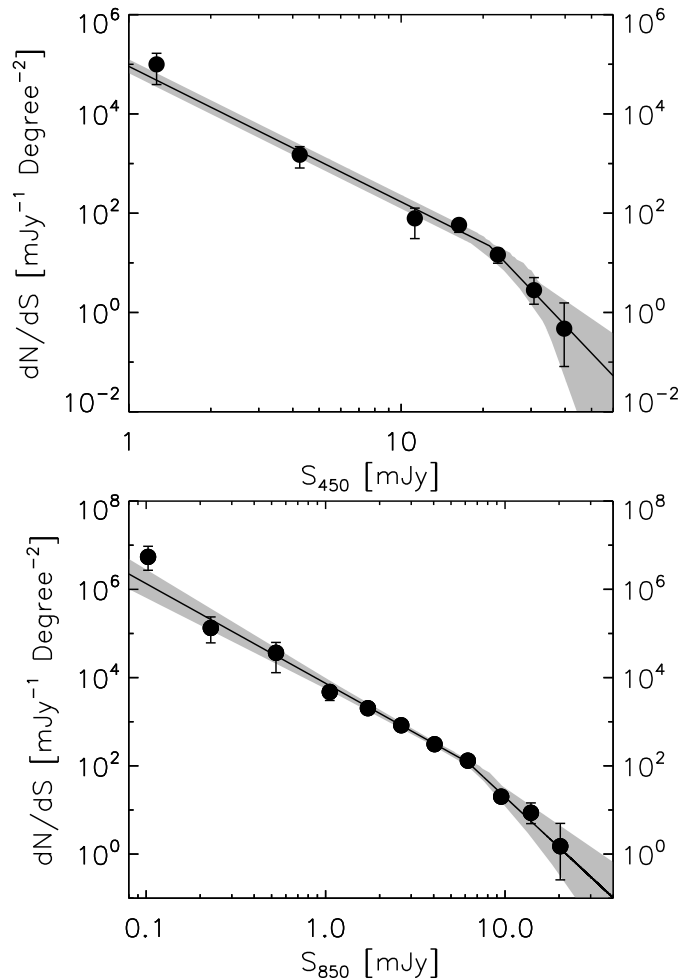}
       \caption{Area-weighted combined counts from all five fields. The best fit broken power law models and the number counts are summarized in Tables \ref{cmcntmdl} and \ref{cmcnt}, respectively.}
     \label{compcnts}
  \end{center}
\end{figure}

Since the discovery of the FIR EBL, resolving the diffuse emission into discrete sources has been one of the primary goals of FIR surveys.  In particular, we would like to determine the flux range of the galaxies that contribute the bulk of the submillimeter light. Thanks to gravitational lensing, our deep SCUBA-2 observations of cluster fields are able to probe to fluxes of $\sim$1\,mJy at 450\,$\mu$m and $\sim$0.1\,mJy at 850\,$\mu$m, an unprecedented depth at 450\,$\mu$m. While showing the counts from each field in the same figure as we did in Figure \ref{allcnts} enables us to visualize the effect of cosmic variance, the amount of EBL resolved based on the model fits to the counts in a single field are poorly constrained ($\sim20-200$\% at 450\,$\mu$m, for example), due to the small number statistics on each field and also cosmic variance. To better constrain the counts models, we combine the counts from all five fields. In each flux bin, we average the corrected, statistically significant differential number counts from each field using an area weighting. We then estimate the Poisson errors using the combined source counts from the contributing fields. We plot the results in Figure \ref{compcnts}. As the figure shows, errors are smaller on the flux bins that are covered by more than one field, and the minimum $\chi^2$ fits on these counts ({\it black curves} with errors in gray shading) are much better constrained than the ones in Figure \ref{allcnts}. The reduced $\chi^2$ are 1.2 and 1.3 at 450 and 850\,$\mu$m, meaning the counts are well described by the models, which are summarized in Table \ref{cmcntmdl}. The area-weighted combined differential counts are given in Table \ref{cmcnt}.

Based on the combined counts models, we measure $113.9^{+49.7}_{-28.4}$($37.3^{+21.1}_{-12.9}$)\,Jy/Degree$^2$ at 450 (850)\,$\mu$m, which corresponds to a full resolution of the 450 and 850\,$\mu$m EBL based on the measurements by \citet{Puget:1996p2082}. However, our measurements also correspond to $80.0^{+34.9}_{-19.9}$ ($85.7^{+48.4}_{-29.6}$)\% of the 450 (850)\,$\mu$m EBL if we adopt the EBL measurements from \citet{Fixsen:1998p2076}. The percentage ranges expand to 48--153\% (44--178\%) if we consider the errors on the EBL measurements. We note that the slope of the faint end counts
results in a weak divergence in both bands so that the counts must flatten at fainter fluxes than those measured here.
The slopes must turn over before 0.5~mJy at 450\,$\mu$m and 0.1~mJy at 850\,$\mu$m in order not to significantly
exceed the measured EBL.

\begin{figure}
 \begin{center}
    \leavevmode
      \includegraphics[scale=0.6]{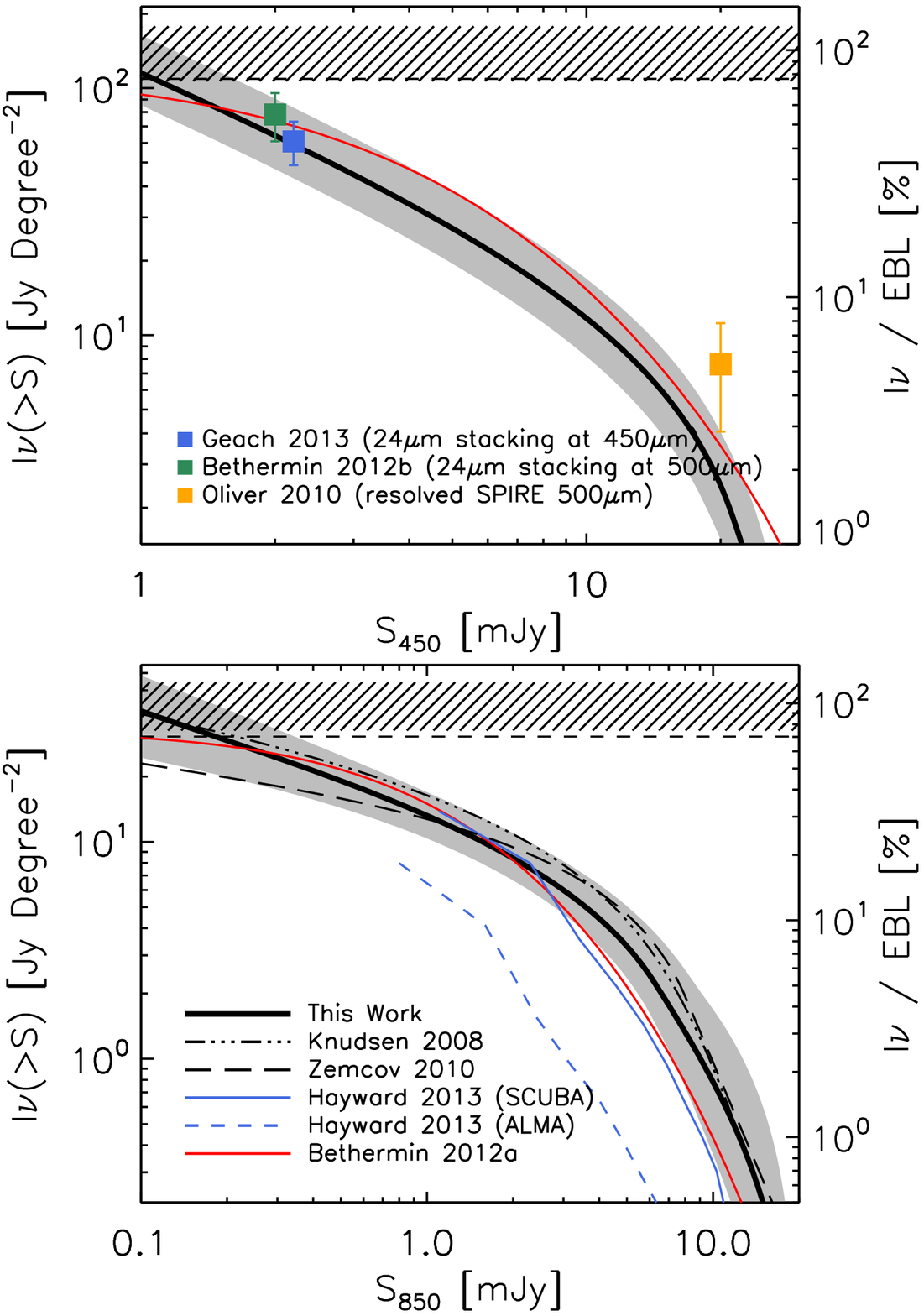}
       \caption{Cumulative extragalactic background light as a function of flux at 450\,$\mu$m ({\it upper}) and 850\,$\mu$m ({\it lower}). The black curves are our best fit broken power law models described in Table \ref{cmcntmdl} with errors in gray shading. The black dashed line \citep{Puget:1996p2082} and the hatched regions \citep{Fixsen:1998p2076} are the background light measured using the COBE satellite. The counts predictions by \citet{Hayward:2013lr} and \citet{Bethermin:2012fk} are plotted as blue and red curves. A few results from the literature on the amount of resolved 450/500\,$\mu$m background light are also shown as colored squares in the upper panel. They are the 24\,$\mu$m sample stacking results on SCUBA-2 450\,$\mu$m maps (blue; \citealt{Geach:2013kx}), 24\,$\mu$m sample stacking results on {\it Herschel} 500\,$\mu$m maps (green; \citealt{Bethermin:2012yq}), and the directly resolved 500\,$\mu$m background light from {\it Herschel} \citep{Oliver:2010p11204}. The dot-dot-dot-dashed \citep{Knudsen:2008p3824} and long-dashed curves \citep{Zemcov:2010uq} represent the counts model curves from SCUBA surveys at 850\,$\mu$m.}
     \label{ebl}
  \end{center}
\end{figure}

We plot the cumulative EBL as a function of flux in Figure \ref{ebl}, where we show that 90\% of the 450\,$\mu$m EBL is contributed by the sources with $S_{450\,\mu m} < 10$\,mJy, and the majority ($\sim$70\%) comes from the sources with 1\,mJy $ < S_{450\,\mu m} < 10$\,mJy. We computed the IR luminosities ($L_{IR}$) for the dominant sources that contribute the bulk of the 450\,$\mu$m EBL by assuming the average redshift ($z=1.3$; \citealt{Geach:2013kx}) and a single temperature modified blackbody spectral energy distribution (SED) with a fixed dust emissivity ($\beta$ = 1.5). The dust temperatures of the SMGs are diverse, ranging from 25$-$50\,K \citep{Magnelli:2012lr}; Thus, we determined all possible $L_{IR}$ values within this dust temperature range for SMGs with 450\,$\mu$m fluxes from 1$-$10\,mJy. We also determined $L_{IR}$ values for the dominant 850\,$\mu$m sources (these have fluxes between 0.1$-$1\,mJy) by assuming $z = 3.0$ \citep{Barger:2012lr,Vieira:2013vn,Hayward:2013lr}.
At both wavelengths, the majority of the dominant sources have $L_{IR}$ between 10$^{11}$ and 10$^{12}$\,$L_\odot$, analogous to local luminous infrared galaxies (LIRGs; \citealt{Sanders:1996p6419}). Note that this result is not sensitive to the assumed redshifts. 
While most SMG studies have focused on the more luminous galaxies that can be detected in single-dish observations of blank fields, it is these lower luminosity galaxies that dominate the submillimeter EBL.

In Figure \ref{ebl} we also show other EBL measurements from the literature at both wavelengths. They all agree nicely with our results, except for the one from the confusion limited {\it Herschel} SPIRE 500\,$\mu$m survey \citep{Oliver:2010p11204}. This is almost certainly a consequence of the fact that the results from \citet{Oliver:2010p11204} are based on a wide area survey where they are able to recover very bright but rare sources that are missed by our observations. However, the contribution of these bright sources to the EBL is small, and correcting for them does not significantly change the calculation of the EBL contribution of the full counts.

We also compare our results with model predictions. The empirical model predictions based on the IR SED templates by \citet{Bethermin:2012fk} are plotted as the red curve in both panels in Figure \ref{ebl}. The predictions match our results very well at 450\,$\mu$m, while the ones at 850\,$\mu$m are slightly lower {at the bright end}. 
While there is no treatment of multiplicity in the model by \citet{Bethermin:2012fk}, the semi-analytical model by \citet{Hayward:2013lr} took this issue into account, and their single-dish 850\,$\mu$m predictions (blue solid curve) agree with our results from $\sim1-2$\,mJy but are low at higher fluxes.

 \citet{Hayward:2013lr} also predicted the counts without any blending (blue dashed curve). At the bright end these are almost an order of magnitude lower than the single-dish results. However, as shown in Section \ref{cdfnsma}, subject to the uncertainty on the multiple fraction, as well as the apparent disagreement between the interferometric counts at 850\,$\mu$m, it is hard to draw a solid conclusion about the true shape of the 850\,$\mu$m counts. In fact, our interferometric counts are statistically indistinguishable from our single-dish counts. Future interferometric follow-up of full submillimeter samples should resolve this issue. 

Interestingly, without considering source blending and by fitting the counts from the mid-infrared to the radio, \citet{Bethermin:2012fk} were able to reasonably model the 450\,$\mu$m counts. This could mean that the majority of the 450\,$\mu$m sources are at $z<3$, in agreement with recent deep SCUBA-2 COSMOS 450\,$\mu$m results by \citet{Geach:2013kx}, \citet{Casey:2013uq} and {\citet{Roseboom:2013lr}}.

\section{Summary}
\label{secsum}

We have presented very deep number counts at 450 and 850\,$\mu$m obtained using the SCUBA-2 submillimeter camera mounted on the JCMT. Based on a mixture of cluster lensing fields (A1689 and A370) and blank fields (CDF-N, CDF-S, and COSMOS), we have determined the number counts over a wide flux range. At 450\,$\mu$m, we detected significant counts down to $\sim$1\,mJy, an unprecedented depth at this wavelength. By integrating the number counts to our flux limits, we find that we have measured $113.9^{+49.7}_{-28.4}$\,Jy/Degree$^2$ of the 450\,$\mu$m EBL, which corresponds to $80.0^{+34.9}_{-19.9}$\% of the EBL measured by \citet{Fixsen:1998p2076} using the COBE satellite. The results show that the majority of the 450\,$\mu$m EBL is contributed by the sources with $S_{450~\mu{\rm m}}$ between 1--10\,mJy, and these sources are likely to be the ones that are analogous to the local luminous infrared galaxies (LIRGs). At 850\,$\mu$m, we resolved $37.3^{+21.1}_{-12.9}$\,Jy/Degree$^2$ of the EBL, corresponding to $85.7^{+48.4}_{-29.6}$\% of the COBE measurement by \citet{Fixsen:1998p2076}. If we consider the large uncertainties on the COBE measurements, the uncertainty of the percentage of the EBL resolved expands to 48--153\% (44--178\%) at 450 (850)\,$\mu$m. Our SCUBA-2 observations revealed statistically different number counts in the CDF-S field compared with the ones obtained by the LABOCA observations, which could be explained by the discrepancy in the number of sources discovered by the two observations. We also showed that there is little field-to-field variance and that source blending (multiplicity) has only a small effect on the shape of the counts determined with SCUBA-2.

\vspace{5mm}
{\it Acknowledgments}
We thank the referee for a careful review to improve the manuscript, Axel Wei{\ss}, Ian Smail and Fabian Walter for giving helpful comments on the LABOCA counts toward the CDF-S region, and {Matthieu B\'{e}thermin for the advice on the model predictions}. We also like to thank Tim Jenness, Per Friberg, David Berry, Douglas Scott and Edward Chapin for helpful discussions on data reduction, JAC support astronomers Iain Coulson, Holy Thomas and Antonio Chrysostomou, and JCMT telescope operators Jim Hoge, William Montgomerie, Jan Wouterloot and Callie McNew. We gratefully acknowledge support from NSF grant AST 0709356 (C.C.C., L.L.C.), the University of Wisconsin Research Committee with funds granted by the Wisconsin Alumni Research Foundation (A.J.B.), the David and Lucile Packard Foundation (A.J.B.), and the National Science Council of Taiwan grant 99-2112-M-001-012-MY3 (W.-H.W.). C.M.C. is generously supported by a Hubble Fellowship provided by Space Telescope Science Institute, grant HST-HF-51268.01-A. This research is also supported by a Grant-In-Aid of Research from the National Academy of Sciences, administered by Sigma Xi, The Scientific Research Society. The James Clerk Maxwell Telescope is operated by the Joint Astronomy Centre on behalf of the Science and Technology Facilities Council of the United Kingdom, the National Research Council of Canada, and (until 31 March 2013) the Netherlands Organisation for Scientific Research. Additional funds for the construction of SCUBA-2 were provided by the Canada Foundation for Innovation. The authors wish to recognize and acknowledge the very significant cultural role and reverence that the summit of Mauna Kea has always had within the indigenous Hawaiian community.  We are most fortunate to have the opportunity to conduct observations from this mountain.


\bibliography{bib}

\end{document}